\documentclass[acmsmall,sigconf, screen]{acmart}
\AtBeginDocument{%
  \providecommand\BibTeX{{%
    \normalfont B\kern-0.5em{\scshape i\kern-0.25em b}\kern-0.8em\TeX}}}

\setcopyright{none}
\copyrightyear{}
\acmYear{}
\acmDOI{}%

\acmISBN{}
\acmBooktitle{} 
\acmConference{}{}
\acmArticle{}

\renewcommand\footnotetextcopyrightpermission[1]{} %
\usepackage{_macros}
\usepackage{tabularx}

\begin{document}

\title[Gamifying XAI]{Gamifying XAI: Enhancing AI Explainability for Non-technical Users through LLM-Powered Narrative Gamifications}

\author{Yuzhe You}
\affiliation{
  \institution{University of Waterloo, School of Computer Science}
  \streetaddress{200 University Ave W}
  \city{Waterloo}
  \state{Ontario}
  \country{Canada}
}
\email{y28you@uwaterloo.ca}

\author{Jian Zhao}
\affiliation{
  \institution{University of Waterloo, School of Computer Science}
  \streetaddress{200 University Ave W}
  \city{Waterloo}
  \state{Ontario}
  \country{Canada}
}
\email{jianzhao@uwaterloo.ca}

\renewcommand{\shortauthors}{Yuzhe You and Jian Zhao}

\begin{abstract}
  Artificial intelligence (AI) has become tightly integrated into modern technology, yet existing exploratory visualizations for explainable AI (XAI) are primarily designed for users with technical expertise. 
  This leaves everyday users, who also regularly interact with AI systems, with limited resources to explore or understand AI technologies they use. 
  We propose a novel framework that enables non-technical users to collect insights by conversing directly with visualization elements via LLM-powered narrative gamifications.  
  We implemented a prototype that utilizes such gamification to facilitate non-technical users' exploration of AI embedding projections. 
  We conducted a comparative study with 10 participants to assess our prototype quantitatively and qualitatively. %
  Our study results indicate that although our prototype effectively enhances non-technical users' AI/XAI knowledge, and users believe they learn more through the gamification feature, it remains inconclusive whether the gamification itself leads to further improvements in understanding. 
  In addition, opinions among participants regarding the framework's engagement are mixed: some believe it enhances their exploration of the visualizations, while others feel it disrupts their workflow.
\end{abstract}

\begin{teaserfigure}
  \centering
  \vspace{-4mm}
  \includegraphics[width=\columnwidth]{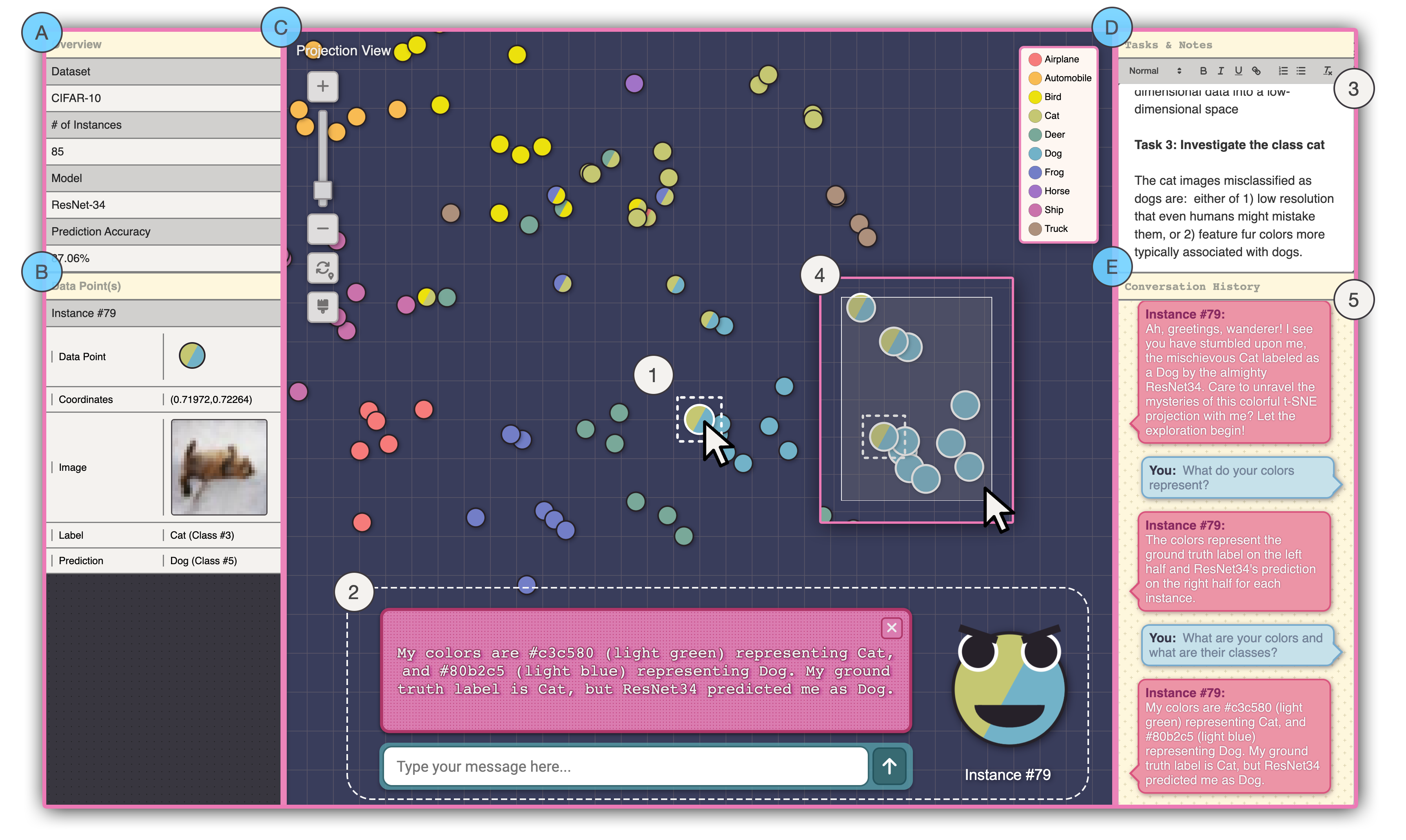}
  \vspace{-8mm}
  \caption{The interface of our prototype integrated with LLM-powered narrative gamification.
  The components include: (a) \textit{Overview View}, a summary of the dataset used and the model; (b) \textit{Data Point(s) View}, which presents instance-level details of all selected data points; (c) \textit{Projection View}, where users can freely explore and converse with the data points and clusters; (d) \textit{Tasks \& Notes View}, where users can record their tasks and collected insights, and (e) \textit{Conversation History View}, where users can revisit their previous conversations with a data point. }
  \label{fig:teaser}
\end{teaserfigure}

\maketitle

\newcommand{\ttest}[2]{{\small $t=#1,p=#2$}}
\newcommand{\pval}[1]{{\small $p=#1$}}

\vspace{-3mm}
\section{Introduction}

As artificial intelligence (AI) becomes closely integrated into modern technology \cite{guo2019survey, kuutti2020survey, tulshan2019survey, lee2021service}, explainable AI (XAI) is increasingly important to bring transparency to all stakeholders, including not only specialists but also everyday users.
Nonetheless, for many, AI remains a complete black box. 
Though exploratory XAI visualizations (\ie, exploration-focused visualizations that explain AI processes and outputs) have been shown to be effective in illustrating the inner workings and performances of AIs (\eg, \cite{cheng2020dece, wang2020cnn, kahng2017cti, kahng2018gan}), they often target users with basic AI knowledge such as machine learning (ML), leaving non-technical users with limited resources. 
To address this, the concept of \textit{gamification} \cite{hallifax2023points}, defined as the application of game-design elements in non-game contexts, emerges as a promising path as it is known to amplify the understanding and engagement of users with little background \cite{caponetto2014gamification, huang2013gamification, mokhtari2021gamified, milosz2020gamification}. 
Nonetheless, while recent efforts have focused on enhancing the interactivity of exploratory XAI visualizations \cite{wang2020cnn, kahng2018gan, ren2016squares, you2023panda}, the potential of incorporating gamification is largely under-explored. 
This presents a gap in the current XAI visualizations, as gamification's capacity to make learning more effective and enjoyable could further enhance their accessibility and effectiveness for a broader audience. 
Among the gamification elements commonly adopted in existing research, ``narrative''-based gamification incorporates storytelling elements and characters into game design, which are known to provide context and guidance within the activity \cite{hallifax2023points}.
Interestingly, recent research and commercial products have shown the versatility and capability of Large Language Models (LLMs) across different domains including narrative-based gamification, one of them being LLM-powered non-playable characters (NPCs) within video games to introduce dynamic and unscripted interactions with players \cite{akoury2023framework, vaudeville2023, inworld2023}. 
This presents a unique opportunity to utilize similar LLM-based narrative gamifications to elevate non-technical users' understanding and experience with XAI visualizations, which we aim to investigate in this work.

Though gamification in XAI visualizations is rather under-explored, some works have tried to investigate gamification in XAI itself. 
For instance, Fulton et al. \cite{fulton2020getting} developed a multi-player game with a purpose (GWAP) where players compete to guess the source input image of a convolutional neural network (CNN) based on its feature visualization images. 
Nonetheless, it is neither a visualization tool nor exploratory, as its purpose is to assess human interpretations of AI explanations.  %
Sevastjanove et al. \cite{sevastjanova2021questioncomb} merged gamification with XAI to create a workspace aimed at engaging users in question classification tasks. 
Yet, the tool is for assisting experts in labeling question types as training data for supervised ML, thus also diverging in focus. 
Other works have explored using LLMs to assist visualization tasks such as supporting automatic creation processes \cite{maddigan2023chat2vis, dibia-2023-lida}, performing Natural Language Interface (NLI) tasks \cite{ko2023natural}, and generating chart captions \cite{ko2023natural, liew2022large}. 
However, these studies do not investigate the potential of LLMs in gamifying visualizations to improve non-technical users' understanding and engagement, which is the primary focus of our research. 

To explore LLM-powered narrative gamification in XAI visualizations and its effects on user comprehension and engagement, we aim to design a novel framework and implement a prototype that integrates such gamification by personifying elements within the visual interface via LLM agents. 
Specifically, we propose a novel form of gamified approach where users can ``speak'' directly to NPC-like visualization elements in natural language to collect insights, turning complex XAI visualizations into a more relatable and engaging narrative.  
Our primary objective is to promote an intuitive understanding of complex XAI visualizations and actively engage non-technical users in the exploration process through gamifications. 
For this work, we primarily focus on how such gamification can enhance users' comprehension and engagement while exploring an interactive t-SNE projection of image-based classifiers' embeddings, a popular form of XAI visualization. 
We choose to use the embedding projection as it is commonly adopted by ML practitioners to visualize high-dimensional data to identify clusters and anomalies to better understand model perceptions and behaviors, as demonstrated by past XAI visualizations such as \cite{tensorflowEmbeddingProjector, kahng2018gan, you2023panda, kahng2017cti}. 
To empirically evaluate our framework, we conducted a between-subjects study with two versions of our prototype, one with and one without the LLM-based narrative gamifications. 
Our study results show that though our prototype effectively enhances non-technical users’ AI/XAI knowledge, and that users believe they learn more through
the gamification feature, it remains inconclusive whether the gamification itself leads to further improvements in understanding. 
Additionally, opinions among participants regarding the framework’s engagement are mixed: some believe the gamification enhances their exploration of the visualizations, while others feel that constantly having to converse with visualization elements disrupts their workflow. 
In summary, we make the following contributions:

\begin{itemize}
    \item We proposed a novel \textbf{framework} that enables non-technical users to collect insights from XAI visualizations by conversing directly with NPC-like visualization elements through LLM-based narrative gamifications. 
    \item Based on our framework, we implemented a \textbf{prototype} that utilizes such gamification to help users explore and understand an embedding projection by interacting with data points or clusters in conversational manners. 
    \item We performed an \textbf{evaluation} with 10 non-technical users to quantitatively and qualitatively assess the effects and usability of our prototype on user comprehension and engagement. 
    We present findings on how its gamifications can help non-technical users better understand and enjoy XAI visualizations and identify areas of improvement and discuss directions for future works. 
\end{itemize}
\vspace{-3mm}
\section{Related Work}

\subsection{Exploratory XAI Visualizations}

In the past, numerous exploratory XAI visualizations have been proposed to help users gain a deeper understanding of the underlying logic and performance of AI models. 
For instance, CNN Explainer \cite{wang2020cnn} visualizes the neuron connections and pathways within a small CNN, enabling users to explore the interplay between its low-level mathematical operations and their high-level model structures. 
GAN Lab \cite{kahng2019does} is designed to promote learning and experimentation with Generative Adversarial Networks (GANs) by allowing users to interactively train GANs using a simple dataset.
The What-If Tool (WIT) \cite{wexler2019if} allows ML practitioners to test the performance of their models in hypothetical situations, and analyze the importance of data features and visualize model behaviors across multiple models. 
ActiVis \cite{kahng2017cti} is a multi-level visualization system for deep neural networks (DNNs) that unifies instance- and subset-level inspection and tightly integrates different views of model performance. 
More examples of exploratory XAI visualizations include Manifold \cite{zhang2018manifold} for DNN comparisons, Squares \cite{ren2016squares} for multi-class model evaluation, RuleMatrix \cite{ming2018rulematrix} for visualizing classifiers with rule-based knowledge representation, and DECE \cite{cheng2020dece} for distilling ML models with counterfactual explanations. 

Nonetheless, despite existing exploratory XAI visualizations proving effective for users with basic AI background in understanding and evaluating AI models, they are heavily techno-centric \cite{ehsan2021expanding} and can be overwhelming for users with no technical backgrounds. 
For instance, CNN Explainer \cite{wang2020cnn} and GAN Lab \cite{kahng2019does}, though being advertised as suitable for non-experts, primarily visualize complex model structures such as neuron activation pathways and layered distributions—concepts that are only comprehensible to those with at least basic ML understanding. 
Moreover, both tools rely on text-based technical terminology (e.g., ``discriminator,'' ``generator,'' ``conv\_1\_1,'' ``max\_pool\_2'') to explain the visualizations, further resulting in understanding barriers for non-technical users. 
Both tools also used observational studies and simple usability questionnaires to demonstrate their effectiveness, lacking concrete study results on how much non-experts were able to ``learn'' through interacting with them. 
The authors of WIT \cite{wexler2019if} initially targeted it at a broad audience including journalists and activists. 
However, after testing with a variety of users, it became apparent that technical prerequisites were necessary to effectively explore the tool. 
More advanced tools \cite{kahng2017cti, zhang2018manifold, ren2016squares, ming2018rulematrix, cheng2020dece} are specifically designed for experienced ML practitioners to evaluate model performance. 
They feature multiple view levels and a high degree of technical complexity, making them unsuitable for those without a technical background.  
Compared to existing tools, our framework introduces a novel narrative-based gamified approach that provides context and guidance for individuals with limited AI background by personifying visualization elements into game-like characters. 
By engaging users in a conversational manner with the visualization, our framework aims to reduce the technical complexity of XAI visualizations, thus enhancing the exploration experience and outcomes for non-technical users.

\vspace{-3mm}
\subsection{LLMs in Visualization}

While the use of LLMs for visualization gamification remains relatively underexplored in the literature, other studies have investigated the potential of LLMs for other visualization tasks. 
For example, LIDA \cite{dibia2023lida} is an LLM-based tool designed for generating grammar-agnostic visualizations, comprising four modules to assist with different stages of visualization generation, including data summarization, visualization goal setting, chart creation, and data-faithful infographic production. 
Similarly, Chat2VIS \cite{maddigan2023chat2vis} adopts a novel end-to-end NL2VIS solution that converts free-form conversational language into visualizations via LLMs. 
VL2NL \cite{ko2023natural} utilizes an LLM-based framework to generate diverse NL datasets from Vega-Lite specifications to streamline the development of Natural Language Interfaces (NLIs) for data visualizations. 
The tool is capable of extracting chart semantics and generating L1/L2 captions with high accuracy, and demonstrates generating and paraphrasing utterances with great diversity. 
Likewise, Liew et al. \cite{liew2022large} explored the use of LLMs for generating compelling captions for data visualizations and reported on the strengths and weaknesses of the GPT-3 model for this task.

Though LLM-based chart generation marks a first step towards leveraging LLMs to enhance visualization interpretation, oftentimes, users' confusion about a visualization system may extend beyond the visualization itself. 
This confusion may also arise from the interface itself or from understanding the technical terminology used within the visualization. 
In such instances, it is natural for users to seek assistance for insight explanations and guidance, a role we believe is also well-suited for LLM agents. 
Furthermore, recent video games have showcased an additional capability of LLMs: the creation of dynamic NPCs capable of generating unscripted responses in real time \cite{akoury2023framework, vaudeville2023, inworld2023}. 
We perceive this as a unique opportunity to leverage LLMs for enhancing XAI visualization through the introduction of narrative gamification elements. 
Unlike simple chatbots, narrative gamification facilitated by LLMs not only could enhances users' understandings, but also offers a uniquely immersive exploration experience. 
By embedding character-like components directly into the visualization, we propose not just to augment users' comprehension but to transform the exploration into a compelling game-like experience. 
We envision that users are likely to engage more deeply with the system due to heightened engagement, thus collecting more insights throughout the exploration process. 
Furthermore, this increased interactivity is expected to encourage users to revisit the system, discovering new insights with each interaction.

\vspace{-3mm}
\subsection{Narrative-based Gamification}

In the context of gamification, ``narrative'' involves incorporating storytelling elements and characters to offer context and guidance outside of traditional game environments \cite{hallifax2023points}. 
Narrative-based elements have been demonstrated to yield multiple positive effects, including enhancing learner memory, motivation, and engagement \cite{bielenberg1997efficacy}, as well as sparking interest in subjects otherwise deemed advanced or uninteresting \cite{papadimitriou2003mythematics}.
For instance, Huynh et al. \cite{huynh2020designing} introduced a role-playing game aimed at promoting visualization literacy in young children by leveraging the presence of narratives in data-related problems involving visualizations. 
Their study results showed that these elements improve engagement without sacrificing learning. 
Langer et al. \cite{langer2014suspenseful} introduced a gamified framework for constructing suspenseful, narrative-based software applications, offering an engaging alternative for inherently complex software. 
This approach was applied to a tutorial for the 3D animation tool Blender, revealing that the narrative provided a sense of hopeful suspense compared to traditional tutorials. 
Hagedorn et al. \cite{hagedorn2019design} applied gameful learning to Massive Open Online Courses (MOOCs) by incorporating a coherent, fully optional detective story that spanned all course weeks. 
After conducting the courses, the authors found that most learners perceived the narrative as either a positive (76\%) element of the course. 
Palomino et al. \cite{palomino2023gamification} developed and validated a Narrative Gamification Framework for Education, which provides educators with tangible guidelines to gamify their lessons by emphasizing the content's gameful transformation rather than the environment. 
The results of their user evaluations and expert feedback collectively demonstrated the effectiveness and potential of their framework in enhancing learner engagement, motivation, and learning outcomes in virtual learning environments.

Inspired by these existing works, we aim to also incorporate narrative-based gamification into our framework, with the goal of enhancing non-technical users' understanding of and engagement with XAI visualizations. 
Specifically, our framework introduces LLM-based NPC-like visualization elements that are ``context-aware'' and capable of providing users with guidance on various aspects of the visualization tool. 
In contrast to traditional narrative gamifications that are static, scripted, and non-exploratory \cite{huynh2020designing, langer2014suspenseful, hagedorn2019design}, our approach is distinguished by its support for a non-linear exploratory process, and is flexible enough to be adapted to different XAI visualization techniques. 
Compared to existing work, our framework aims to further enhance the educational value and engagement of exploratory XAI visualizations by increasing their replay value and diversifying the insights that can be gathered, based on each user's unique background and preferences. 
This allows different users to gain their own unique insights during exploration, and even encourages the same user to potentially discover new insights during subsequent interactions with the system.

\vspace{-5mm}
\subsection{Gamification in XAI}

Though research on gamification within exploratory XAI visualizations remains limited, existing studies have explored the integration of gamification elements into XAI itself. 
For instance, Fulton et al. \cite{fulton2020getting} integrated XAI with Games with a Purpose (GWAP) by developing a multiplayer game in which two players compete to identify the source input of a CNN based on its feature visualization images. 
QuestionComb \cite{sevastjanova2021questioncomb} is an interactive labeling interface for question-type ML data that employs a mixed visual analytics technique, combining concepts from both gamification (e.g., user guidance, feedback and rewards) and XAI. 
Geleta et al. \cite{geleta2023maestro} introduced Maestro, a gamified platform designed to help learners study robust AI by offering goal-based scenarios that immerse students in AI-related assignments within a competitive programming environment. 

Nonetheless, these works diverge in focus from ours and, as such, do not adequately address our research question. 
For instance, the GWAP developed by Fulton et al. \cite{fulton2020getting} is neither a visualization tool nor designed for exploratory purposes, as its primary focus is to understand how humans select and interpret feature visualizations of CNNs, rather than to enhance user interpretation of XAI visualizations. 
QuestionComb \cite{sevastjanova2021questioncomb} is specifically designed for assisting experts in labeling question types for ML training data, and is thus irrelevant to our focus on exploratory XAI visualizations. 
Similarly, Maestro \cite{geleta2023maestro} is an educational platform that is unrelated to XAI visualization but instead offers gamified, AI-based assignments and coursework to students, thus again diverging in focus from our work. 
In reviewing existing gamification work within XAI, we have identified an underexplored area in current research: using gamification to demystify complex XAI visualizations and provide context and guidance for non-technical users. 
This presents a unique opportunity to enhance understanding and interpretation of these visualizations.
Addressing this unexplored opportunity in our study, we aim to contribute to the XAI landscape by presenting a gamified framework designed to mitigate the techno-centric nature of current XAI visualizations. 
By facilitating direct interaction between users and NPC-like visualization elements, our framework seeks to enable users to derive insights more intuitively.

\vspace{-3mm}
\section{Approach}

In this section, we provide a detailed description of our gamified framework's design and the technical strategy implemented for its development.

\vspace{-2mm}
\subsection{Framework Design Rationale}

Due to the complexity of XAI visualizations, non-technical users may find themselves puzzled by many aspects of the visualization system, including:

\begin{itemize}
    \item 1) Functionalities and usage of the interface view components (\textbf{G1});
    \item 2) Technical terms and concepts mentioned in the visualization system that are relevant to AI/XAI (\textbf{G2});
    \item 3) Meanings of visual encodings employed and how to interpret the actual visualizations (\textbf{G3}).
\end{itemize}

In these instances, it is natural for users to want to seek assistance for insight explanations and guidance, a task well-suited for LLM agents. 
Using LLM-powered NPCs as an inspiration, our proposed interaction method ``personifies'' visualization elements into dynamic game-like characters, adding narrative elements (\eg, context, characters, guidance) to the visualization and transforming it into a gamified experience. 
Compared to traditional gamifications that use scripted dialogues/contents \cite{huynh2020designing, geleta2023maestro}, our LLM-based design also maintains a non-linear exploratory process that is flexible to be adapted to different XAI visualizations. 
Here, we first present a scenario featuring our prototype of an interactive 2D t-SNE projection of ResNet-34's CIFAR-10 embeddings to demonstrate the general approach of our gamified framework. 

Suppose a user with limited technical AI/XAI knowledge is exploring the projection of ResNet-34's model embeddings within the visualization interface of our prototype. 
Upon launching the system, they notice that the visualization interface is segmented into several view components. 
A quick overview of the information presented in each view component leads the user to find the interface generally intuitive, except for the central view labeled as ``\textit{Projection View} (\autoref{fig:teaser}-C),'' which appears as a scatterplot filled with colorful points. 
Confused about the purpose and functionalities of this view, the user decides to first list ``Understand how to use Projection View'' as their initial task in the \textit{Tasks \& Notes View} (\autoref{fig:teaser}-D). 
Then, the user decides to click on one of the data points, labeled as instance \#38, to initiate a conversation.
Consequently, an input textbox, along with the data point's avatar—depicted as a shy, blushing character—and a dialogue box, are displayed at the bottom of the interface. 
Instance \#38 briefly introduces itself, suggesting to the user, \qt{I-if you have any questions or need help understanding the projection, feel free to ask!} 
The user begins by typing their query into the input textbox, inquiring instance \#38 for an overview of the Projection View. 
Instance \#38 elaborates, \qt{In the p-projection view, you can see where all the different d-data points are placed based on how s-similar they are to each other according to ResNet-34.} 
The user further asks for clarification on the functionalities of the buttons located at the top left of the Projection View. 
Instance \#38 responds, detailing, \qt{The buttons help you to z-zoom in, out, or reset the view its o-original position. You can also use the brush feature to select multiple data points for further exploration!}
Through this exchange and subsequent follow-up questions, the user gains a comprehensive understanding of the Projection View's purpose and features. 
This process effectively simplifies the initial learning phase, significantly flattening the learning curve associated with adopting the tool (\textbf{G1}).

Nonetheless, while the user has gained an understanding of the view components within the interface, their exploration process is still impeded due to their unfamiliarity with technical terms such as ``ResNet-34,'' ``t-SNE,'' and ``CIFAR-10 dataset.'' 
To familiarize themselves with these AI/XAI terms and better grasp the visualization, the user notes this down as their current task and engages with instance \#76 by initiating another conversation. 
Instance \#76 is characterized as slightly annoyed by the user's presence, adding a unique dynamic to the learning experience. 
\qt{Ugh, what do you want now? [...] This t-SNE thingy shows our positions based on how similar we are. Got it?} 
Instance \#76 begins the conversation with a characterized introduction, effectively engaging the user. 
Confused about the term ``t-SNE,'' the user inquires Instance \#76 for a brief overview of the t-SNE projection concept. 
Instance \#76 replies: \qt{t-SNE [...] is a technique used to visualize high-dimensional data in a lower-dimensional space, making it easier to see patterns and clusters in the data.} 
The user then asks about the meaning of ``dimension'' within the AI context, to which Instance \#76 responds: \qt{Dimension refers to the number of attributes or features used to represent each data point [...].} 
Through engaging in similar exchanges about various technical terminologies, the user develops a thorough comprehension of the terms utilized within each view of the interface  (\textbf{G2}). 
This knowledge equips them to more effectively explore and interpret the visualization system. 
For example, the user now knows that the projection is achieved with ``t-SNE,'' a method for reducing dimensionality for data analysis. 
They also recognize that the AI model visualized is ``ResNet-34,'' an image classifier categorized as a type of CNN. Furthermore, they learn that the image dataset visualized is identified as the ``CIFAR-10 dataset,'' comprising low-resolution images from ten different classes of animals and objects.

Ready to explore the t-SNE projection of data embeddings, the user starts by familiarizing themselves with the visual encodings. 
They observe two categories of data points in the scatterplots: uniformly colored data points, and data points split into two halves, each half colored differently. 
The user opts to click on instance \#79, depicted as a circle with its left half green and the right half blue (\autoref{fig:teaser}-1). 
Curious about its color encoding, the user asks the data point about the significance of its colors (\autoref{fig:teaser}-2). 
Instance \#79 explains, \qt{The colors represent the ground truth label on the left and ResNet34's prediction on the right. My green left half means I am a cat, while my blue right half means ResNet-34 predicted me as a Dog. }
Knowing what the colors stand for now, the user proceeds to examine the hues of all data points and clusters within the projection. 
They observe that ResNet-34 appears to struggle with inter-class discrepancy in accuracy, particularly within the class ``cat'' (\textbf{G3}). 
More precisely, the user notices that ResNet-34 often misclassifies images of other subjects as cats and incorrectly identifies several cat images as dogs.
To understand the reasons behind its suboptimal class fairness, they note ``Investigate the class cat'' as their current exploration task (\autoref{fig:teaser}-3) and identify several data points of cat within the dog cluster. 
Intrigued to learn more about this cluster, the user activates the brush toggle button, enabling them to select and interact with multiple data points simultaneously.
The user drags their cursor to create a selection box around the cluster, thus highlighting every point predicted as a dog (\autoref{fig:teaser}-4). 
Subsequently, a new avatar emerges, this time symbolizing the entirety of the dog cluster, and the \textit{Data Point(s) View} (\autoref{fig:teaser}-B) refreshes to exhibit instance-level details and attributes of all points within the cluster. 
The user begins by inquiring about the prediction accuracy within cluster, in which the cluster responds, \qt{In our cluster of 11 instances, 8 were predicted correctly. The common misclassifications were instances of Cats being incorrectly predicted as Dogs.} 
The user then asks about the potential reasons for the misclassifications, to which the cluster responds: \qt{[...] It could be due to similarities in features like fur texture or body shapes that make them more challenging to differentiate.} 
This inspires the user to navigate to the Data Point(s) View to examine the actual images within the cluster. 
They observe that the misclassified cat images are either of such low resolution that even humans might mistake them for dogs, or feature cats with fur colors more typically associated with dogs. 
This similarity in critical visual features poses a challenge for ResNet-34 in accurately classifying these images. 
The user documents their insights in the Tasks \& Notes View regarding the model's frequent confusion between cats and dogs. 
As they gather more information from conversations, they record significant insights under their noted task, similar to how a video game player collects insights and takes notes while interacting with NPCs (\textbf{G3}). 
The conversational history with each data point is saved within the \textit{Conversation History View} (\autoref{fig:teaser}-E), enabling the user to revisit it in the future and easily recall all insights (\autoref{fig:teaser}-5).

\begin{figure*}[!t]
\centering
  \includegraphics[width=0.8\linewidth]{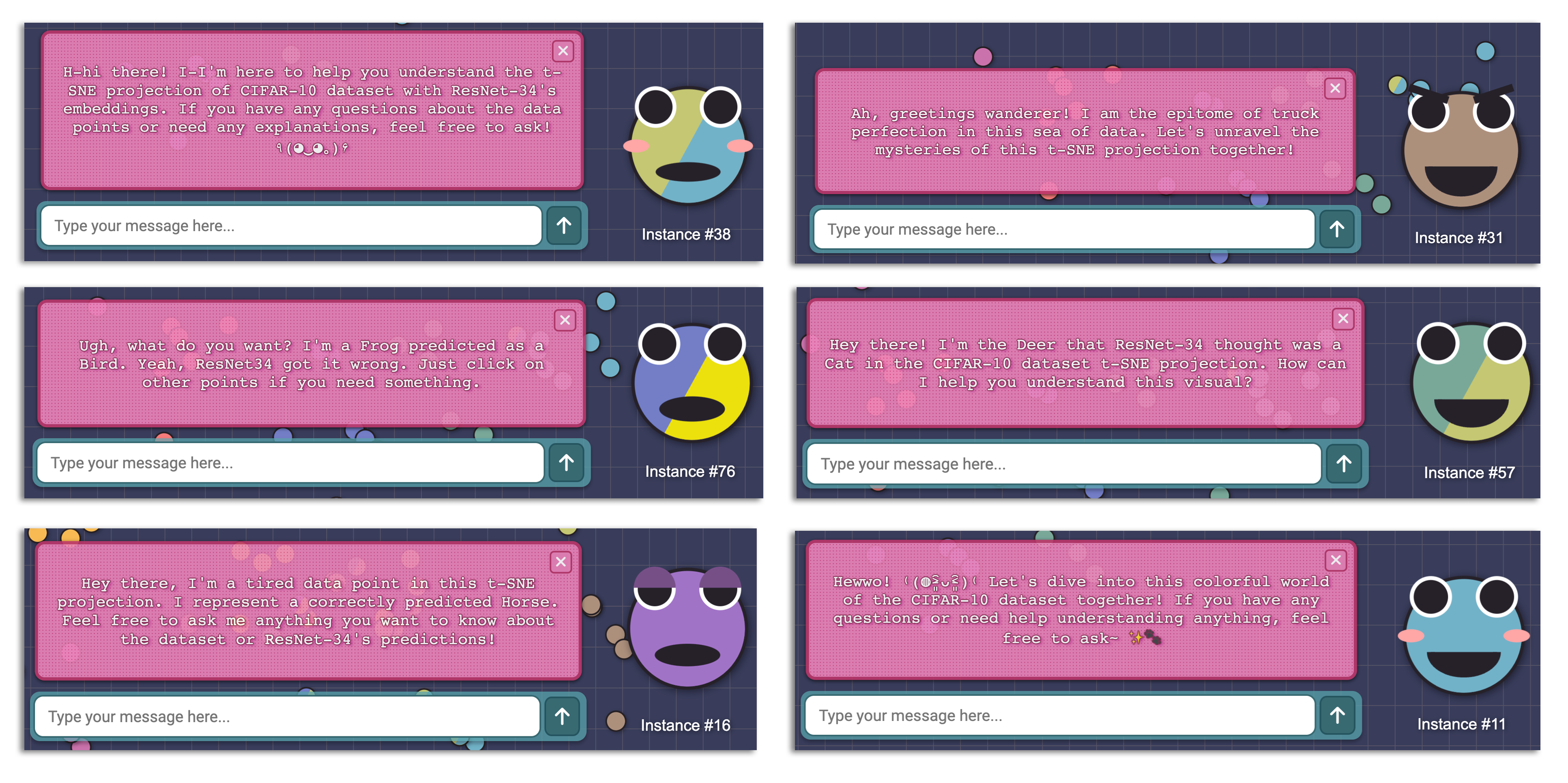}
  \vspace{-5mm}
\caption{Examples of different personalities we incorporated into our data points.}
  \label{fig:personalities}
   \vspace{-5mm}
\end{figure*}

\vspace{-2mm}
\subsection{Prototype Implementation}

For our prototype, we developed a web application using Node.js. 
The frontend visualizations were created with D3.js to construct the interface components of the visualization tool. 
Our design was partly inspired by the What-If Tool (WIT) \cite{wexler2019if}.
The components we included are: (a) \textit{Overview View} (\autoref{fig:teaser}-A), a summary of the dataset used and the model's overall performance; (b) \textit{Data Point(s) View} (\autoref{fig:teaser}-B), which presents detailed information and attributes of all highlighted data points; (c) \textit{Projection View} (\autoref{fig:teaser}-C), the central view in which users can freely explore and interact with the t-SNE distribution of image representations, as well as selecting data points and clusters to initiate conversations with; (d) \textit{Tasks \& Notes View} (\autoref{fig:teaser}-D), where users can note down their exploration goals and collected insights, and (e) \textit{Conversation History View} (\autoref{fig:teaser}-E), where users can revisit their previous conversations with a visualization element.

To integrate LLM capabilities into our prototype, we utilized OpenAI's API, specifically the GPT-3.5-turbo model, to enable dynamic conversational interactions. 
Communication between our application and OpenAI's services was achieved with an HTTP client library, with server-side logic designed to construct queries based on user interactions and send them to OpenAI's API to retrieve AI-generated responses. 
Each data point was personified as a unique, game-like character by assigning it a distinct personality from a predefined array of personalities detailed in natural language (\autoref{fig:personalities}). 
Additionally, to ensure "context-awareness," we provided each data point with a prompt encompassing the visualization interface's functional aspects, their guiding role for users, the visualized models and datasets, essential dataset and model statistics (\eg, overall accuracy, class distribution, color representations), and the specific attributes of selected data points (\eg, prediction labels, scatterplot locations). 
This approach enabled the simulation of meaningful conversations.
Moreover, we leveraged the ElevenLabs API, a text-to-speech service, allowing the data points to audibly communicate their GPT model-generated responses.
\vspace{-3mm}
\section{Experiment and Results}

To evaluate how our framework can help non-technical users with their explorations of XAI visualizations, we conducted a comparative study with participants possessing limited technical background in AI/ML and minimal familiarity with CNNs or t-SNE projections. 
Our investigation focuses on two key aspects of our gamified approach: (\textbf{A1}) the effectiveness of our framework in improving users' understanding and insight gathering of XAI visualizations, and (\textbf{A2}) the impact of gamification on enhancing user engagement with the visualization tool. 

\subsection{Study Setup}

\textbf{Participants and Apparatus.}
We recruited 10 participants (P1 $\sim$ P10; six men, four women; age 22 $\sim$ 33) with limited technical background in AI/ML. 
They came from different backgrounds such as Civil Engineering, Computer Science, Industrial Engineering, and more. 
A demographic and screening questionnaire was distributed to recruit participants, including questions that asked them to self-rate their technical experience with AI and ML.
Specifically, on a 7-point Likert scale (self-rated; 1=``Novice'', 7=``Expert''), we recruited participants that satisfied all the following constraints: AI experience $\leq$ 5, ML experience $\leq$ 4, familiarity with CNN models $\leq$ 2, and familiarity with t-SNE projections $\leq$ 2. 
Their median experience with AI is 2 (\iqr{1}), their median ML experience is 2 (\iqr{0.75}), their median CNN familiarity is 1 (\iqr{1}), and their median t-SNE familiarity is 1 (\iqr{0}). 
The study was conducted remotely via Zoom. 

\textbf{Task and Procedure.}
For our controlled experiment, we utilized a between-subjects study design, in which we compared our gamified prototype $P_1$ with a baseline variant $P_0$ that lacks the LLM-powered narrative gamification. 
Participants were divided into two groups to compare the effects of our interventions: one group experienced the baseline condition, and the other experienced the gamified condition. 
Participants assigned to $P_0$ received an additional PDF document that included tutorials on the system interface and explanations of key AI/XAI concepts.
We assign them in such a way that the average level of ML and AI experience was comparable between the two conditions. 
Specifically, the participants assigned to the baseline condition had an average AI experience of 2.8 (\sd{1.303}), and an average ML experience of 2 (\sd{1.225}). 
Meanwhile, the particiapnts assigned to the gamified condition had an average AI experience of 2.6 (\sd{0.894}), and an average ML experience of 1.8 (\sd{0.837}). 
We divided our study into five phases: pre-quiz, interaction (task completion), post-questionnaire, post-quiz, and interview. 
To initiate the study, we first required participants to take a pre-quiz consisting of eight questions to assess their general understanding of AI/XAI concepts.
They were given 10 minutes to complete this pre-quiz. 
After introducing the assigned prototype to the participants, we explained to the participants the tasks they were expected to complete during their interaction. 
They were provided with an online form requiring them to complete 12 tasks related to 1) understanding the system's interface, 2) grasping AI/XAI concepts, and 3) decoding visual encodings and interpreting the visualization. 
Participants had 40 minutes to complete these tasks and were encouraged to use the tasks \& notes view (\autoref{fig:teaser}-D) for documenting any notable insights about the visualization. 
To ensure authentic responses, participants were instructed to explain their answers in their own words, preventing direct copying and pasting of responses generated by LLMs. 
After the interaction, participants were asked to complete a Likert scale post-questionnaire that included NASA-TLX, which included six questions on the cognitive demand of task completion with their prototype. 
Additionally, there were 13 other questions that focused on the learning effects and usability of the system. 
Following the post-questionnaire, participants were asked to take another post-quiz, which consists of 21 questions across three knowledge categories they were expected to learn during their interaction: 1) interface, 2) AI/XAI concepts, and 3) visualization interpretation. 
Participants had 20 minutes to complete this post-quiz.
To conclude each study session, an interview is conducted with each participant to gather additional qualitative data. 
This involves asking for their opinions on the learning effects and usability of their assigned prototype. 
For those assigned to the gamified prototype, specific questions about the gamification feature were also asked to understand their detailed views on our gamified framework. 

\vspace{-3mm}
\subsection{Results and Analysis: Task Performance}

\textbf{Quiz performance within groups.}
Within each group, we first analyzed participants' quiz performances on the eight AI/XAI concepts questions before and after the intervention. 
Through paired t-tests, we evaluated the learning gains within each condition, and found that both the baseline (\ttest{-3.763}{0.019}) and experimental groups (\ttest{-3.505}{0.024}) demonstrated statistically significant improvements. 
These results suggest that both systems are effective in enhancing users’ understanding of AI/XAI concepts, regardless of the inclusion of the gamification feature.

\textbf{Quiz performance and task completion between groups.}
To assess the difference in participants' performance across the conditions, we analyzed their average scores in quiz performance and task completion. 
Additionally, we utilized unpaired t-tests to statistically evaluate the differences between the groups. 
When examining the pre-quiz scores, participants in the gamified condition, on average, outperformed those in the baseline condition (\mean{68\% > 60\%}). 
However, the unpaired t-test indicated that this difference is not statistically significant (\ttest{-0.524}{0.616}), suggesting that participants in the gamified condition did not have a higher level of AI/ML experience. 
Additionally, participants in the gamified condition achieved a high task completion rate of 100\% (\sd{0}), compared to the baseline group’s completion rate of 83.07\% (\sd{0.206}). 
The unpaired t-test indicates that this difference is marginally statistically significant (\ttest{-0.1833}{0.141}). 
Despite the unpaired t-tests not revealing significant differences between the two groups, possibly due to the small sample size utilized by this study, participants from the gamified condition generally outperformed those in the baseline. 
This advantage was observed in task completion accuracy (\mean{84.61\% > 75.38\%}; \ttest{-0.739}{0.498}), scores from the post-quiz questions on AI/XAI concepts (\mean{97.5\% > 92.5\%}; \ttest{-0.600}{0.565}), and the overall post-quiz performance (\mean{91.43\% > 88.57\%}; \ttest{-0.894}{0.406}). 
The detailed results are presented in \autoref{table:results}. 

\begin{table*}[htb!]
\small
\centering
\renewcommand{\arraystretch}{1}
\begin{tabularx}{\textwidth}{>{\bfseries}X l >{\raggedright\arraybackslash}X X}
\toprule
\textbf{Metric} & \textbf{Baseline Average} & \textbf{Experimental Average} & \textbf{Unpaired t-tests} \\
\midrule
Pre-quiz & 60\% (\sd{0.185}) & 68\% (\sd{0.244}) & \ttest{-0.524}{0.616} \\
\midrule
Task completion rate & 83.07\% (\sd{0.206}) & 100\% (\sd{0}) & \ttest{-1.833}{0.141} \\
\midrule
Task completion accuracy & 75.38\% (\sd{0.274}) & 84.61\% (\sd{0.054}) & \ttest{-0.739}{0.498} \\
\midrule
Post-quiz (AI/XAI concepts) & 92.5\% (\sd{0.112}) & 97.5\% (\sd{0.056}) & \ttest{-0.600}{0.565} \\
\midrule
Post-quiz & 88.57\% (\sd{0.080}) & 91.43\% (\sd{0.070}) & \ttest{-0.894}{0.406} \\
\midrule
\bottomrule
\end{tabularx}
\vspace{1mm}
\caption{The results of quiz averages, task completion rate/accuracy, and unpaired t-tests from participants across both conditions. }
\vspace{-5mm}
\label{table:results}
\end{table*}

\vspace{-3mm}
\subsection{Results and Analysis: Perceived Cognitive Load (NASA-TLX)}
We employed the NASA-TLX to measure the perceived workload associated with each prototype. 
Both systems share similar mental (\md{5} for both conditions; \pval{0.515}) and physical demands (\md{2} for both conditions; \pval{0.912}), suggesting that the inclusion of gamification feature does not impose additional mental or physical strain on users. 
Compared to baseline, the gamified system has lower temporal demand (\md{5 < 6}; \pval{0.5219}), leading to better performance (\md{3 < 5}; \pval{0.1363}) and less effort (\md{4 < 5}; \pval{0.1931}) with marginal statistical significance, as well as less frustration (\md{3 < 5}; \pval{0.6684}). 
The overall perceived workload, measured by averaging all six raw NASA-TLX scores, was also lower for the gamified system than for the baseline (\md{3.833 < 4.667}; \pval{0.5296}). 
Consequently, participants generally found the prototype integrated with our gamified framework to be less cognitively demanding compared to the baseline, although this difference was not statistically significant. 

\begin{figure*}[!t]
\centering
  \includegraphics[width=\linewidth]{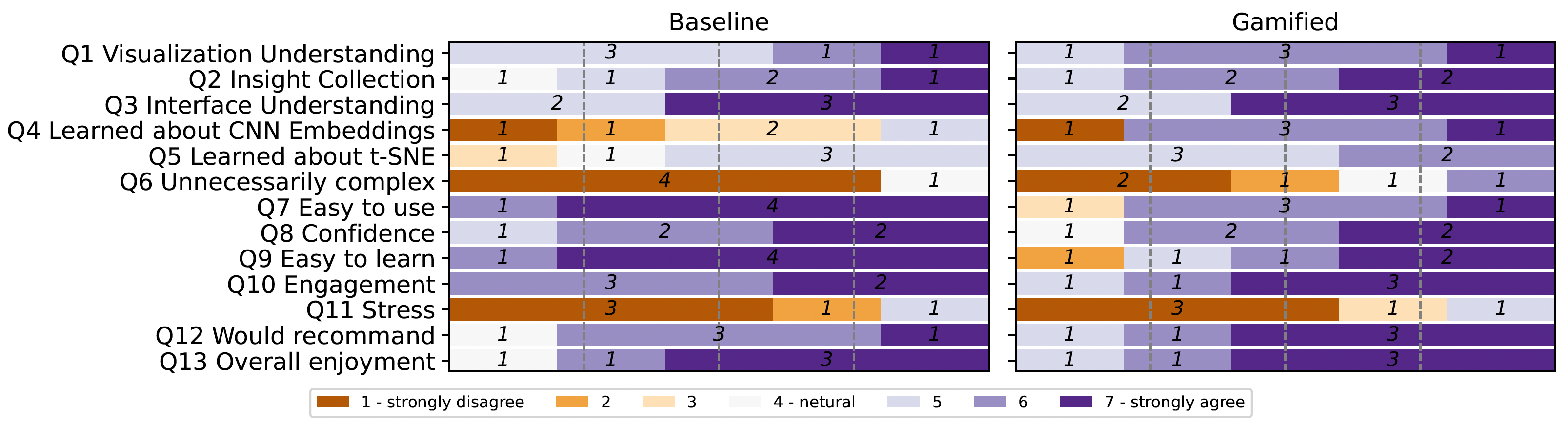}
  \vspace{-5mm}
\caption{User perceptions of the learning effects and usability of both the baseline and the gamified prototypes, measured using self-defined, seven-point Likert scales.}
  \label{fig:likert}
   \vspace{-5mm}
\end{figure*}

\subsection{Results and Analysis: Perceived Learning Effects and Usability}
Additionally, we asked the participants to self-rate their perceptions of the learning effects and usability of their assigned prototypes, as detailed in \autoref{fig:likert}. 
Generally, participants from the gamified condition reported a better understanding of visualizations and AI/XAI concepts compared to those from the baseline. 
This was reflected in their perceptions of the amount of visualization understanding they gained (Q1: \md{6 > 4}; \pval{0.433}), how much they felt they learned about CNN embeddings (Q4: \md{6 > 3}; \pval{0.110}), and their perceived learning of t-SNE (Q5: \md{5} for both; \pval{0.076}), which exhibits a marginally significant p-value. 
This could be attributed to the gamification intuitively allowing users to seek explanations for concepts and visualizations that are unclear to them. 
For instance, while participants assigned to the baseline claimed that they could grasp the meanings of the positions and colors of the data points, they felt that the system offered only surface-level explanations an failed to provide in-depth insights into why certain data points were misclassified or positioned as they were within the projection. 
P9 (baseline) explained, \qt{I wish the [baseline] offered deeper explanations for its classifications and why it clusters data points the way it does; this would help me further understand what's happening in the visualization.}
P8 (baseline) made a similar comment, \qt{I felt the [baseline] only scratched the surface of understanding, particularly with the mechanisms behind image classification and embeddings, leaving me wanting more in-depth explanations.} 
On the other hand, participants assigned to the gamified prototype commended its gamification feature for enhancing their understanding of the XAI visualization, attributing this improvement to the feature's ability to simplify the system's technical complexity. 
P1 (gamified) mentioned, \qt{I like how if I don't understand anything, I can simply ask questions in plain English without having to search these things up myself.}
P4 (gamified) agreed, stating that \qt{Using natural language with images made it easier for me to grasp and interact with what's behind the data points; I could learn further with follow-up questions for more detailed answers.}

When examining participants' ratings on their interface understanding (Q3), no significant difference was observed between the two groups, as ratings from both were notably high (\md{7} for both conditions; \pval{1.0}). 
During interviews, participants explained that the prototype's interface was already straightforward and intuitive, and that those assigned to the gamified condition also did not find it necessary to rely on the gamification for assistance in navigating the interface. 
In the future, we could potentially extend our gamification framework to XAI visualizations with more complex interfaces to better assess whether it impacts users' interface comprehension compared to systems without gamification. 

Nonetheless, compared to the baseline, participants generally found the gamified prototype to be relatively more complex and difficult to use. 
This was reflected in their perceptions of whether the system was unnecessarily complex (Q6: \md{2 > 1}; \pval{0.287}), its ease of use (Q7: \md{6 < 7}; \pval{0.083}), and its ease of learning (Q9: \md{6 < 7}; \pval{0.193}). 
Overall, the baseline system received higher ratings for its intuitiveness. 
Some participants from the gamified condition expressed that they felt it was unnecessary for users to have to hear visualization elements ``speak'' or to engage in conversations with them, as this disrupted their actual workflow while exploring the visualizations. 
Certain participants also found the gamified system more challenging to use due to the inherent inaccuracies of the GPT model used for the data points, as they noticed that sometimes these LLM agents provided answers that were clearly incorrect, leading to a loss of trust. 
P7 (gamified) commented, \qt{Sometimes the data points give me incorrect answers for things I know, thus making me question the accuracy of the information provided.} 
P1 (gamified), on the other hand, expressed frustration that data points sometimes respond to their follow-up questions using more technical terms, without elaborating on the areas they hoped would be further clarified, thus making the system less intuitive to use. 

In terms of engagement, although the gamified prototype received a slightly higher rating than the baseline (Q10: \md{7 > 6}; \pval{0.908}), participants assigned to the gamified group expressed mixed feelings. 
This ambivalence largely arises from the issues previously discussed, such as interruptions to the exploration workflow from needing to constantly interact with data points, frustration with the system due to inaccuracies (\ie, model hallucinations), and the excessive use of technical terms in explanations. 
Nonetheless, during the interviews, most participants still expressed positive sentiment towards the engagement of our gamification framework and highlighted the beneficial impacts it could have on non-technical users’ exploration experience. 
P2 (gamified) explained, \qt{I love the feature of conversing with data points for its simplicity in obtaining answers and the charm of each point's unique personality, making the interaction engaging and informative.}
P4 (gamified) stated, \qt{The system's gamified approach significantly boosts my motivation to study; it sparks my curiosity, encouraging me to explore further, revisit data points, and easily understand the classification through color differentiation, inspiring a continuous loop of discovery and learning.}

\vspace{-3mm}
\section{Discussion}

In this section, we discuss key considerations and limitations of our existing gamification framework for exploratory XAI visualizations, offering insights and directions for future research. 

\vspace{-3mm}
\subsection{Key considerations}

In discussing the integration of narrative gamification within XAI visualizations, our study highlights several key considerations. 
Firstly, our gamified framework has shown potential in motivating users to engage more deeply with the visualizations, sparking curiosity and encouraging a more proactive approach to learning. 
This suggests that our framework can inspire users to explore beyond the surface level of provided content.
Moreover, this could also potentially enhance the ``replay'' value of XAI visualizations, motivating users to revisit the tool for further exploration and potentially uncover new insights. 
This not only deepens the understanding of complex visualizations but also ensures sustained engagement with the educational tool, reinforcing learning through repeated interaction. 
Secondly, another notable observation from our study is the comfort participants experienced while interacting with the data points within our framework. 
Many participants shared during interviews that they felt they could ask any question without the fear of sounding uninformed or facing judgment. 
This not only demonstrates the potential of our gamified approach in creating more engaging and approachable XAI visualization tools but also suggests it could significantly lower the barriers to AI learning. 
By promoting a more open and supportive exploration environment, our approach might encourage a wider audience of non-technical individuals to deepen their understanding of AI.

\vspace{-3mm}
\subsection{Limitations}

Despite our framework showing potential in enhancing user understanding and engagement, it still suffers from limitations.
A notable challenge identified in our study is the users' distrust of the system, which arises from the limitation of the system providing incorrect answers or using overly technical language. 
This highlights the need to balance between accuracy and accessibility in gamification design for XAI visualizations. 
To mitigate distrust and improve user experience, we need to also take into account the clarity and accuracy of LLM responses, ensuring that the system supports non-technical users in their learning process without causing confusion or misinformation. 
Another limitation was the potential disruption of users' exploration workflows due to the need to constantly interact with data points, as brought up by our participants. 
While this frequent engagement is designed to facilitate learning, it may also interrupt the natural visualization exploration process. 
Addressing this limitation would require a careful design consideration to ensure that the gamification elements are seamlessly integrated into the system to support the user experience rather than hindering it. 
Another straightforward solution is to make the gamification feature optional, allowing users to toggle it based on their preference.

\vspace{-3mm}
\section{Future work}

While our study highlights the potential of our gamification framework, there is still significant room for further exploration and enhancement.
Firstly, our gamified approach was tested only on one popular type of XAI visualization, specifically the t-SNE embedding projection. 
To thoroughly assess whether our framework is effective across different XAI visualizations, future work should aim to incorporate the gamified framework into a broader range of XAI visualization techniques, such as neuron connection visualization, and DNN saliency map. 
This extension would provide a more comprehensive understanding of the framework's generalizability.

Additionally, our statistical analyses, including unpaired t-tests and Mann-Whitney tests, did not demonstrate significance in many cases, which may be attributed to the small sample size of our study. 
With only ten participants, divided evenly across conditions, the power to detect meaningful differences was likely limited. 
To truly validate the effectiveness and usability of our framework, future studies must be conducted with a larger participant pool. 
This would enhance the reliability of our statistical findings and provide a more solid foundation for evaluating the proposed gamified framework's contributions to enhancing the exploration of XAI visualizations.

\vspace{-3mm}
\section{Conclusion}

In this study, we propose a novel gamification framework that enables non-technical users to understand XAI visualizations and gather insights by engaging in conversations directly with NPC-like visualization elements through LLM-powered narrative gamifications. 
Based on our framework, we implemented a prototype that utilizes the gamification to enhance non-technical users' exploration of an interactive t-SNE embedding projection. 
We conducted a between-subjects study to assess our framework quantitatively and qualitatively, and our results show that our gamified prototype effectively enhances users' AI/XAI knowledge, and users also believed that they were able to learn more from the gamificiation feature. 
Nonetheless, it remains inconclusive whether the gamification feature itself contributes to further improvements in understanding, and participants' opinions on the feature's engagement were mixed. 

\bibliographystyle{ACM-Reference-Format}
\bibliography{main}

\end{document}